\documentclass[conference]{IEEEtran}
\usepackage{xcolor,soul,framed} 
\usepackage{minted}

\colorlet{shadecolor}{yellow}
\usepackage[pdftex]{graphicx}
\graphicspath{{../pdf/}{../jpeg/}}
\DeclareGraphicsExtensions{.pdf,.jpeg,.png}

\usepackage[cmex10]{amsmath}
\usepackage{array}
\usepackage{mdwmath}
\usepackage{mdwtab}
\usepackage{eqparbox}
\usepackage{url}
\usepackage{subfigure}
\usepackage{tabularx}
\usepackage{tikz}
 
\usepackage{caption}

\ifCLASSINFOpdf
\else
\fi
\usepackage{fancyhdr}
\begin{document}
%
\title{Demo: BeGREEN Intelligence Plane for AI-Driven O-RAN Energy Efficiency}

\author{\IEEEauthorblockN{Miguel Catalan-Cid, Jorge Pueyo}
        \IEEEauthorblockA{
            \textit{i2CAT Foundation, Spain}\\
            \{miguel.catalan, jorge.pueyo\}@i2cat.net
        }
\\
\IEEEauthorblockN{Jesus Gutierrez}
\IEEEauthorblockA{
            \textit{IHP - Leibniz-Institut für innovative Mikroelektronik, Germany }\\
teran@ihp-microelectronics.com}
\and
\IEEEauthorblockN{Juan Sanchez-Gonzalez}
\IEEEauthorblockA{
            \textit{Universitat Politècnica de Catalunya, Spain}\\
            juansanchez@tsc.upc.edu 
        }
\\
\IEEEauthorblockN{Mir Ghoraishi }
\IEEEauthorblockA{
            \textit{Gigasys Solutions Ltd, UK}\\
mir@gigasys.co.uk }
}

\maketitle
\thispagestyle{fancy} 

\begin{abstract}
Cellular networks are undergoing a revolutionary transform with the advent of O-RAN architectures and AI/ML solutions. O-RAN's Non-Real-Time and Near-Real Time RAN Intelligent Controllers open the door to the implementation of automated control-loops that can provide RAN optimisations in numerous scenarios and use cases, and which can be further empowered by AI-driven approaches. Energetic sustainability has raised as one of the main optimisations targets due to the impact of mobile networks on global energy consumption. To this end, the BeGREEN project aims at enhancing the energy efficiency of beyond 5G networks by defining novel AI/ML-based methods at RAN and edge infrastructure. This paper presents BeGREEN Intelligent Plane, a novel framework which implements and exposes AI/ML workflows to O-RAN-based optimisations targeting energy efficiency. We also describe an exemplary application of the Intelligent Plane and its AI Engine, which aims at providing AI-driven cell on/off control.  
\end{abstract}


%
\IEEEpeerreviewmaketitle

\section{Introduction}\label{sec:intro}
The transition from 5G to beyond 5G (B5G) and 6G mobile communication networks brings a paradigm shift not only in terms of enhanced performance and increased connectivity, but also in addressing critical issues related to the environmental implications associated to a higher energy consumption. Improving the planning, deployment, and management of B5G and 6G networks is imperative to counteract the rising energy consumption trend. Overcoming these challenges require innovative architectural revisions and novel algorithmic solutions to promote sustainability and mitigate the environmental impact of cellular networks \cite{greenran}. Notably, the Radio Access Network (RAN) consumes more than 70\% of the total energy of a 5G system, making its optimisation a top priority.

The consolidation of the O-RAN architecture, which advocates for disaggregated, virtualized and software-based components, connected through open and standardised interfaces, entails a significant opportunity to intelligently manage the RAN with the aim of improving the network performance, and reduce energy consumption \cite{oranenergy}. Particularly, the Non-Real-Time RAN Intelligent Controller (non-RT RIC), and Near-Real Time RAN Intelligent Controller (Near-RT RIC) provide the required functionalities to develop and host the so-called rApps and xApps implementing, respectively, long-term and almost real-time optimisations through automated control-loops.

Furthermore, the integration of Artificial Intelligence and Machine Learning (AI/ML) introduces a cognitive layer that can learn from historical data, adapt to evolving network dynamics and make adequate decisions for improving the network performance and the energy efficiency \cite{aigreen}. The concrete specification of the supported AI/ML workflows in the O-RAN is still on-going \cite{oranaiml}. Nevertheless, it will allow several options for providing AI/ML workflow services, for example model management, model training, model inference, data preparation, etc., at the Service Management and Orchestatration (SMO), the Non-RT RIC, the Near-RT RIC or through external components. Tightly (image-based) and loosely (file-based) coupled approaches will be also supported, allowing rApps/xApps to host the models and the training/inference runtimes or to use exposed AI/ML services provided by other components, respectively. 

In this context, besides the user plane and  data plane, BeGREEN introduces an Intelligent Plane, which introduces AI/ML control and management plane functions to reduce the overall energy consumption of the RAN infrastructure \cite{d21}. The proposed Intelligent Plane incorporates an AI Engine, which will provide a serverless execution environment hosting the AI/ML models, offering inference and training services to the rApps/xApps by following a loosely coupled approach.  

The rest of the paper is organised as follows. Section II provides an overview of the related work in the context of open-source RIC implementations to implement automated control-loops. Section III presents the architecture of the BeGREEN Intelligent Plane and the AI Engine, with focus on the designed AI/ML workflows. Section IV presents the main components and workflows involved in a specific use case based on energy-efficient cell on/off control. Finally, conclusions are summarised in section V.

\section{Related work}\label{sec:related}
This section briefly presents relevant open-source RIC implementations and their utilisation to implement intelligent and automated control loops. The O-RAN Alliance and the Linux Foundation are collaborating by means of the O-RAN Software Community (OSC) to develop open-source Non-RT and Near-RT RIC solutions aligned with O-RAN specifications. In parallel, two additional initiatives, the Open Air Interface (OAI) Alliance and the Open Network Foundation (ONF), which also collaborate with the O-RAN Alliance, are developing their own open-source RIC solutions. 

The development of the Non-RT RIC by the OSC \cite{osc} relies on the publication of cumulative releases, each of them covering different components, interfaces and workflows defined by the O-RAN Alliance specifications. While the first releases were mainly centered on the communication with the Near-RT RIC through the A1 interface, including the A1 Policy controller (A1-P) and the exposure of enrichment information (A1-EI), the actual focus relies on the R1 interface to provide Data Management and Exposure (DME) and Service Management and Exposure (SME) services. DME services are implemented through the Information Coordinator Service (ICS) \cite{ics}, which serves as a data subscription platform designed to streamline the interaction between data producers and consumers. Regarding SME services, a 3GPP Common API Framework (CAPIF) approach is being developed and evaluated in the last releases, being its definition still in early stages. Additionally, last releases have incorporated an rApp Manager to support rApp life-cycle management (LCM). 

The OSC also provides an implementation of the Near-RT RIC on a microservices architecture, serving as a host for independently developed xApps \cite{oscnear}. Serving as a mediator, the Near-RT RIC platform facilitates interactions between xApps and RAN elements via E2 interface and with network operators through A1 and O1 interfaces. Key components, including xApps lifecycle management, configuration management, and security, are integral to the platform's functionality. Several relevant projects are built on the OSC's Near-RT RIC, such as OpenRAN Gym \cite{openrangym}, which implements E2 towards Open Air Interface (OAI) gNBs or NS-3 simulator, or Open AI Cellular (OAIC) \cite{oaic}, which interfaces srsRAN-based gNBs. 

FlexRIC \cite{flexric} is the proposed framework from OAI to implement a programmable near-RT RIC. It interfaces with the OAI radio stack via the O-RAN-defined E2-interface for real-time monitoring and control of the RAN, although it claims to be vendor-agnostic. It also introduces a novel interface, called E42, to support interaction between the xAPPs SDK and the near-RT RIC. Additionally, FlexRIC defines internal applications (iApps) which allow the development of specialized and low-latency controllers for common control operations, such as RAN slicing control, and whose operations are exposed to xApps to implement intelligent control-loops. 

The ONF has also developed its own near-RT RIC, called SD-RAN \cite{sdran}, compliant with the O-RAN architecture and built on other ONF platforms like Open Network Operating System (ONOS). It is designed around the use of different micro-services with very delineated roles and responsibilities, such as all the necessary terminations for the main standard O-RAN interfaces, i.e., E2, A1 and O1. The SD-RAN-in-a-Box (RiaB) solution, provides a SD-RAN cluster able to operate within a single host machine by deploying the whole SD-RAN infrastructure on Kubernetes: the ONOS RIC, an Open Mobile Evolved Core (OMEC) and the RAN. Regarding the RAN, it is compliant with OAI solutions, and also provides a RAN simulator, which allows simulating several RAN Centralized Units (CU), Distributed Units (DU) and Radio Units (RU) via the O-RAN E2AP. Additionally, it also provides several built-in xApps, like traffic steering or RAN slice management. 

Regarding the integration with AI/ML services, both FlexRIC and SD-RAN frameworks provide the means to develop AI/ML-driven optimisations through xApps, though actually they don't implement a specific solution. For instance, SD-RAN is being leveraged by ONF's SMART-5G project in order to develop AI/ML-driven energy savings solutions for mobile networks \cite{smart5g}. On the other hand, OSC has started on the implementation of an AI/ML Framework exposed to Non-RT and Near-RT RICs \cite{oranaimlfw}. It implements training services through a training host platform based on Kuberflow pipelines, while model serving is performed inside each rApp or xApp using Kserve through a tightly couple approach. However, this implementation is still on its early stages, and, as mentioned in the introduction, O-RAN specification will also support loosely coupled solutions. This latter is the approach being selected by BeGREEN in order to implement its Intelligent Plane and AI Engine, as will be described in the next section. 
\section{BeGREEN Intelligent Plane and AI Engine}\label{sec:ip}
\begin{figure*}[ht]
  \centering
  \captionsetup{justification=centering}
  \includegraphics[width=1.5\columnwidth]{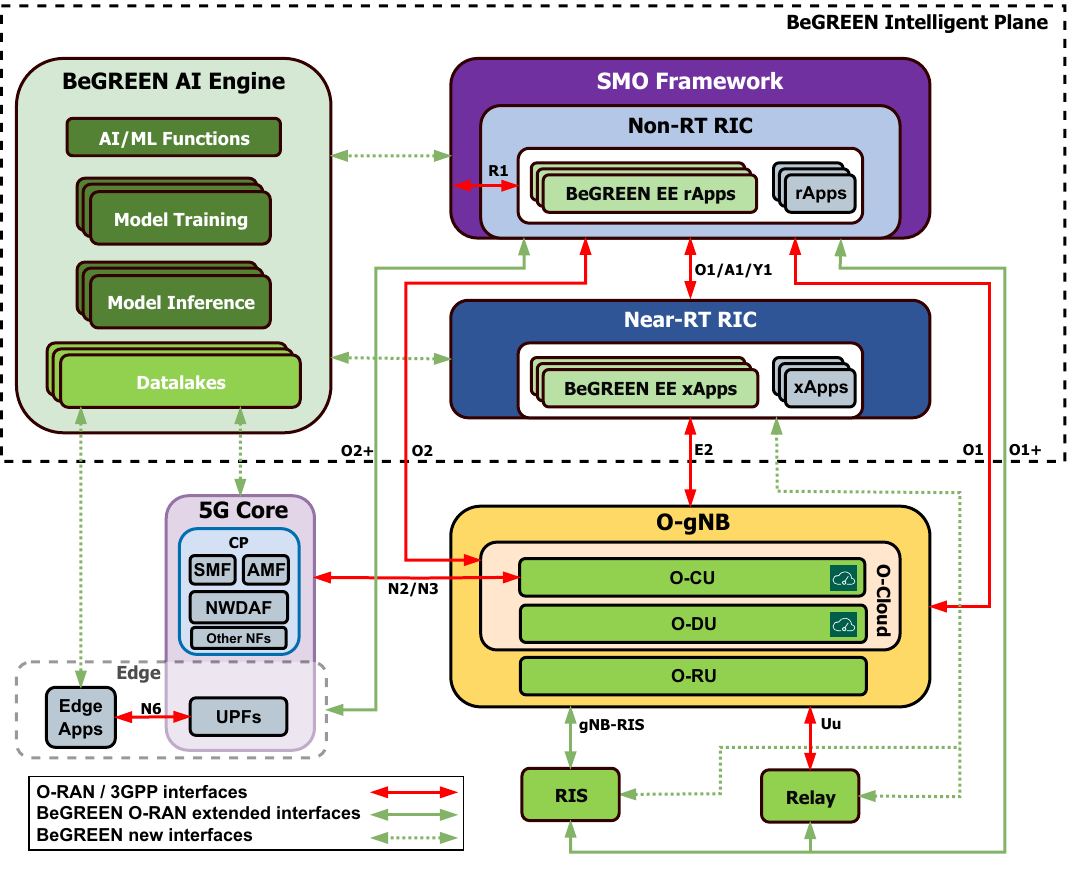}
  \caption{BeGREEN architecture including the Intelligent Plane}
  \label{fig:int_plane}
\end{figure*}

BeGREEN targets energy efficient optimisations at RAN and Edge infrastructure domains. Therefore, the Intelligent Plane, whose main architecture is illustrated in Figure \ref{fig:int_plane}, can be seen as a cross-domain management function \cite{crossdomain}, integrating monitoring, analytics and control operations from Edge, Core and RAN domains. This enables the development of feature-rich ML models, hosted in the AI Engine, which can be exposed to analytics consumers such as rApps and xApps to implement energy efficient automated control-loops. Furthermore, the Intelligent Plane aims at integrating the control of Reconfigurable Intelligent Surfaces (RIS), fixed relays or Relay User Equipments (i.e. UEs with relaying capabilities) \cite{d21}, which are technologies currently beyond the O-RAN's scope. Therefore, in addition to O-RAN/3GPP compliant interfaces, Figure \ref{fig:int_plane} shows proposed extensions, denoted as O1+ and O2+, and new interfaces to monitor and control these elements, and to integrate the AI Engine. 

The objective of the O1+ interface is to enable Non-RT control and monitoring of RIS and relay elements, in a similar way to compliant O-Nodes. For instance, in the case of the relays, this will enable the collection of network measurements and performance indicators that can be useful to identify regions with coverage problems, geographical space/time traffic distributions in the network, etc. Then, this data could be exploited by AI/ML-based rApps to take adequate relay control decisions, such as activation/deactivation or reconfiguration. In the case of O2+, it will leverage O2 functionalities to monitor and manage the resources of the Edge infrastructure hosting software-based user-plane functions such as the 5G Core UPF, with the objective of reducing energy consumption without impairing traffic performance. The final specification of these interfaces is still under definition in the BeGREEN project and it is out of the scope of this paper. 

Regarding the the Non-RT and Near-RT RICs, two specific implementations are considered within BeGREEN project. On the one hand, in the case of the Non-RT RIC, we are leveraging the implementation from OSC, focusing on the exposure of ML models through the R1 interface and the ICS component, which implements DME services as introduced in Section \ref{sec:related}. Operating as an intermediary, the ICS simplifies the relationship between data producers and data consumers by establishing data subscriptions, referred to as Information Jobs, decoupled from the specific data generator. Furthermore, these Information Jobs have the flexibility to draw data from multiple sources, enhancing versatility in data consumption within the Intelligent Plane. On the other hand, the Near-RT RIC is based on a commercial cloud-native solution developed by a partner of the consortium. Nevertheless, as will be further described in this section, the objective of the followed loosely coupled approach is to facilitate the integration of AI/ML workflows through the AI Engine and the associated Assist rApps/xApps, independently of the RIC implementation.

\begin{figure*}[ht]
  \centering
  \captionsetup{justification=centering}
  \includegraphics[width=1.5\columnwidth]{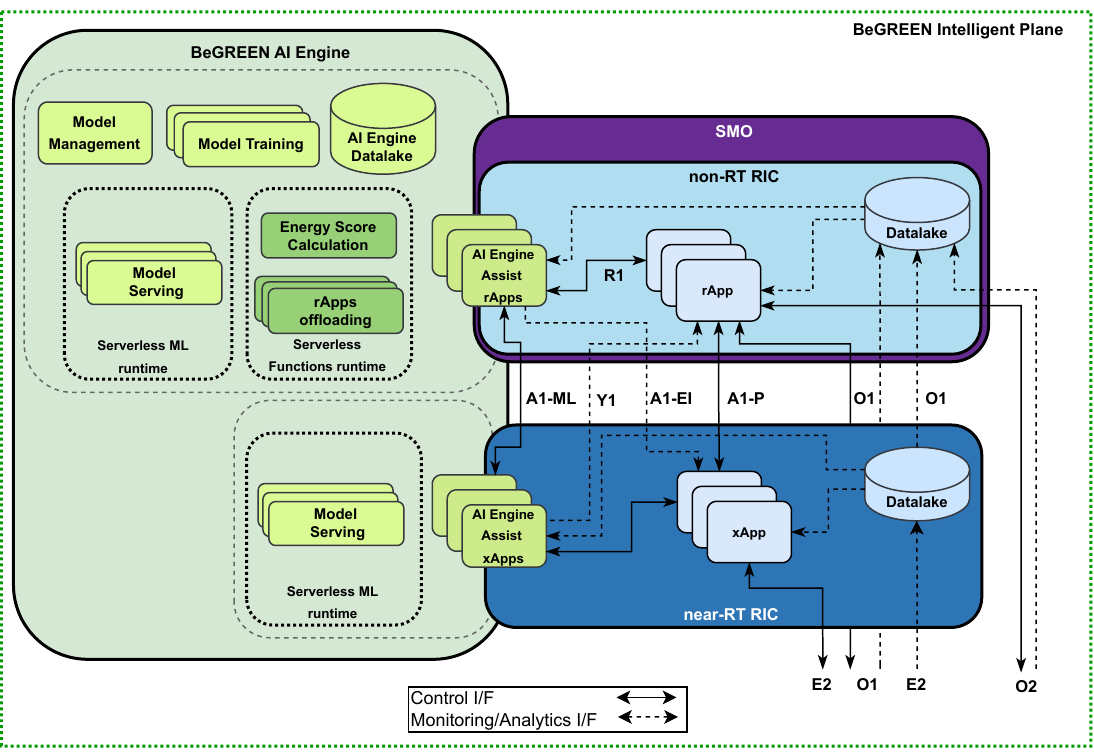}
  \caption{BeGREEN AI Engine architecture}
  \label{fig:aiengine}
\end{figure*}

Completing the Intelligent Plane architecture, the AI Engine will host the ML models to offload workload from the RICs, but also implement the required AI/ML workflows or services. As detailed in Figure \ref{fig:aiengine}, it will include model management, monitoring, training, serving and the datalake with prepared data. In addition to ML models, the AI engine will host other functions which may be used intensively by the rApps/xApps, such as the BeGREEN Energy Score calculation \cite{d21}. This Key Performance Indicator (KPI) will be used to characterize the energy efficiency of the network and its components, and of the applied optimisations. It will also help to detect areas or components with low efficiency, triggering the orchestration of the required optimisations. 

The proposed AI Engine adopts a loosely coupled approach, wherein AI/ML models are hosted within the AI Engine rather than being embedded in the control rApps/xApps that require their outputs. This approach allows for the independent management of ML models by dedicated rApps/xApps. Notably, any control rApp/xApp can access the ML model outputs, which are exposed as  data types (e.g., offering load predictions for specific cells), promoting model reuse. Moreover, this design permits the deployment of ML models for training or inference on servers or clusters separate from the RICs, enabling offloading through serverless computing and hardware acceleration.

To make the AI/ML workflow services of the AI Engine accessible to the rApps/xApps, the BeGREEN project introduces the concept of AI Engine Assist rApps/xApps. These specific rApps/xApps are linked to an ML model, exposing its outputs to other rApps/xApps by acting as data producers through ICS coordination. They also facilitate communication between the AI Engine and the RICs for procedures like ML monitoring or retraining. The communication between these rApps/xApps and the AI Engine relies on the definition of a common interface or Software Development Kit (SDK), while each is individually responsible for its ML model requirements, such as necessary input parameters, data for training or inference, triggered pipelines, and more. Initially we are considering four primary AI/ML workflows, which are described as follows:
\begin{itemize}
\item \textbf{ML Model Creation and Training}: To create and train an ML model, the model developer acquires the necessary data from the AI Engine datalake by engaging with the RICs. This may involve deploying rApps or xApps to produce and expose the required data. After analyzing the data, the developer generates and trains the model. Once trained, the model is published in the catalog, and the developer establishes the inference pipeline, optionally incorporating monitoring and retraining pipelines. Finally, the developer creates the AI Engine Assist rApp/xApp, which interfaces with the AI Engine through the AI Engine-RIC interface and serves as an ML model producer.
\item \textbf{ML Model Inference (Non-RT domain)}: For ML model inference at the Non-RT RIC level, a rApp developer or deployer creates or deploys a control rApp subscribing to the ML model outputs exposed by the AI Engine Assist rApp through the R1 interface. The AI Engine Assist rApp generates inference data through the serverless on-demand deployment of the ML model's inference pipeline in the AI Engine. The control rApp receives the ML output and generates an action, policy or data through O1 (O-Nodes), O2 (O-Cloud), A1 (Near-RT RIC, xApps) or R1 (rApps) interfaces. Alternatively, an xApp developer or deployer can create or deploy a control xApp subscribing to the ML model outputs through the A1-EI interface. In this case, the control xApp obtains the predictions and generates actions through the E2 interface (E2-Nodes).
\item \textbf{ML Model Inference (Near-RT domain)}: In this case, an xApp developer or deployer creates or deploys a control xApp subscribing to the ML model outputs exposed by the AI Engine Assist xApp through the Near-RT RIC. The AI Engine Assist xApp acquires inference data from the ML model's inference pipeline in the AI Engine, deployed either in the same server or in a nearby server (e.g., on the same edge resources) to ensure Near-RT runtime. Finally, the xApp receives the ML model output and generates an action through the E2 interface. Alternatively, inference outputs may be exposed northbound (e.g., to the RIC rApps) or to external components (e.g., application servers) through the Y1 interface.
\item \textbf{ML Model Monitoring and Retraining}: The AI Engine Assist rApp triggers the retraining of an ML model based on the monitoring pipeline. Retraining may use prepared data from the RICs, obtained through the AI Engine Assist rApp itself. In some scenarios, an A1-ML interface between Assist rApps and xApps could be used to exchange ML model information. For instance, different  Assist xApps deployed in different near-RT RICs but associated to the same model could report model performance, triggering model monitoring and retraining by the Assist rApp. Similarly, in Federated Learning scenarios, this interface may be used to exchange model weights. 
\end{itemize}

\begin{figure*}[ht]
  \centering
  \captionsetup{justification=centering}
  \includegraphics[width=1.9\columnwidth]{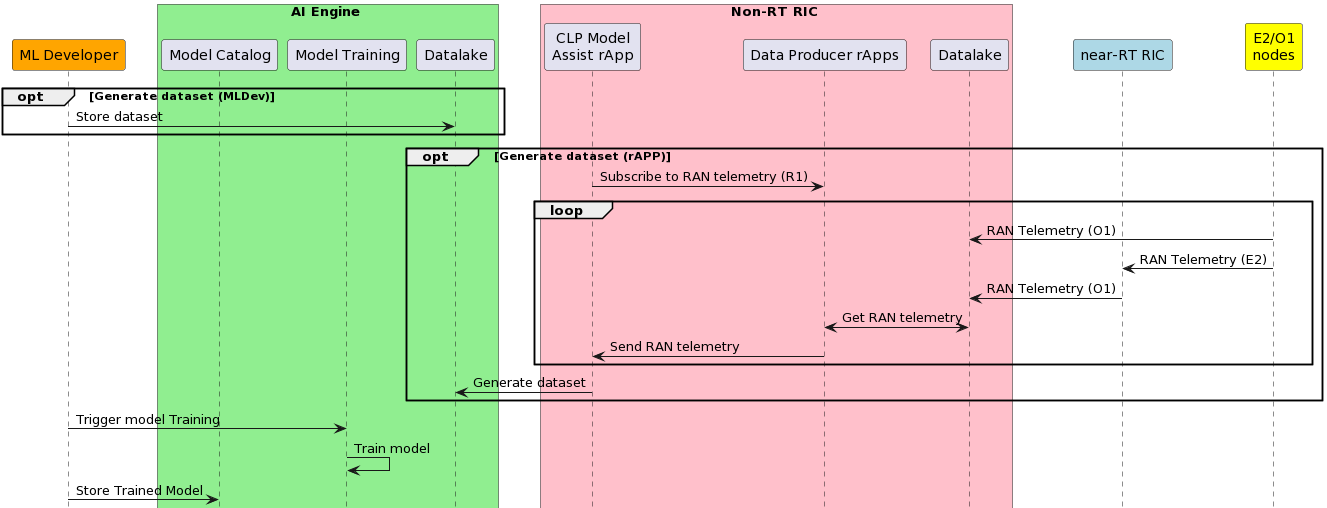}
  \caption{Cell on/off switching use case: Model training phase}
  \label{fig:training}
\end{figure*}

To implement these workflows, certain components will leverage existing open-source frameworks or dedicated solutions in the realms of MLOps. MLOps has emerged as a pivotal solution in addressing the intricate challenges associated with training, deploying, monitoring, and sustaining ML models within production environments. The primary objective of MLOps is to streamline the entire lifecycle of an ML project, with a focus on not only enhancing the efficiency of model development but also ensuring the effective deployment and management of these models. Therefore, MLOps frameworks are characterized by their emphasis on automation, scalability, and reproducibility, which is why most of them are integrated with orchestration platforms such as Kubernetes. Particularly, we have selected the MLRun framework, which covers the whole ML pipeline, including their automation and monitoring. MLRun also allows serverless automation through Nuclio, which is a high-performance serverless framework that supports execution over CPUs and GPUs.

\section{Use Case: Energy efficient RU control}\label{sec:usecase}

One of the main energy efficient optimisations targeted by BeGREEN aims at intelligently controlling RUs through switching on/off their cells based on traffic status and predictions. According to this, cells that are expected to serve very low traffic in certain periods of time can be switched off in order to reduce energy consumption, consequently steering their traffic among active neighbor cells. This is one of the principal energy saving uses cases introduced by O-RAN Alliance in its report \cite{oranenergy}, due to the significant impact of the RU operations on the overall energy consumption of the network. In this context, AI/ML plays a pivotal role, as predictions related to load and energy consumption become key to enable the implementation of proactive automated control loops, contributing to energy efficiency without impairing the status of network traffic \cite{aigreen}.

\begin{figure*}[ht]
  \centering
  \captionsetup{justification=centering}
  \includegraphics[width=1.9\columnwidth]{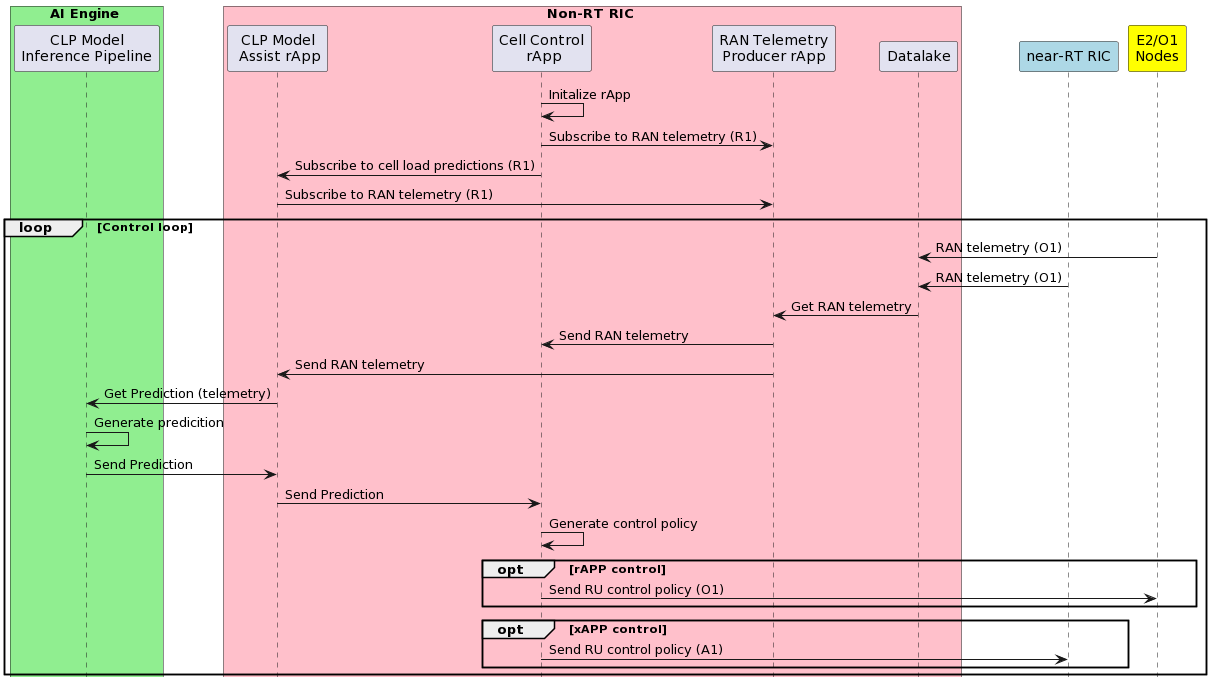}
  \caption{Cell on/off switching use case: Model serving and control phase}
  \label{fig:inference}
\end{figure*}

In this section we illustrate the main Intelligent Plane components and workflows involved in this use case. The solution requires the development of three essential artifacts. First, the Cell Load Predictor (CLP), that will offer the predicted traffic per cell. Secondly, the CLP Model Assist rApp, which will manage the required ML workflows through the AI Engine and the exposure of model outputs to other rApp through the R1 interface. Finally, a Cell Control rApp, which will consume the predictions and perform the cell control according to them. 

Regarding the CLP model, we have access to a real dataset from a mobile network operator in Spain. We are currently evaluating regression algorithms such as Gradient Boosting through the XGBoost Python library, which is widely recognized for its high accuracy and scalability, particularly in handling large datasets \cite{xgboost}. Additionally, it can be used for time series forecasting, which seems especially valuable since we found in the dataset an evident correlation between load, energy consumption, and time trends. Therefore, we are also exploring the integration of ML models to provide energy consumption predictions according to load inputs.

Figure \ref{fig:training} illustrates the workflows required during the training phase of the model. Two main options are contemplated. In the first one, the model is trained by an offline dataset already present in the AI Engine, which could comprehend data from real deployments captured in different measurement campaigns. Alternatively, for instance for retraining purposes, the Assist rApp may be be used to generate, augment or update the dataset by subscribing, consuming and preparing data obtained from the O1 and/or E2 nodes. The required data will be produced by other rApps and exposed through the R1 interface, as implemented by the ICS component. Once the dataset is ready, the model can be trained and stored. Note the monitoring and retraining could be triggered by the Assist rApp, according to its implementation logic.  

Once trained, the CLP model inference will be exposed to the control rApps through the Assist rApp performing as data producer. This way, any rApp requiring this data will need only to create a subscription through the R1 interface and the ICS component. Figure \ref{fig:inference} illustrates the case of a Cell Control rApp, which subscribes to RAN telemetry (RAN Telemetry Producer rApp) and cell load predictions (CLP Model Assist rApp) in order to perform decisions. Note that in this phase, the Assist rApp may also require to consume data to generate the inference (e.g., the current load of the cell). 

After generating the subscriptions, the automated control-loop performed by the Cell Control rApp will consist of obtaining the data and the prediction, deciding the RU control action, and performing the action through O1 (direct control to the O-Node) or through A1 (indirect control through Near-RT RIC or xApp policies). Moreover, the Assist rApp has the capability to transmit performance KPIs regarding the inference. These KPIs can then be leveraged by the RU Control rApp to enhance its decision-making process. For instance, during peak-hour periods, decisions to deactivate cells might adopt a conservative approach based on the accuracy of predictions.

\section{Conclusions}
This paper presented BeGREEN approach to provide AI/ML services to O-RAN RICs through the implementation of an Intelligent Plane. The AI Engine works an external component which implements AI/ML workflows through a MLOps framework, making use of Assist rApps/xApps to manage and expose the ML models in a loosely coupled approach. Exploiting the DME features of the R1 interface, ML model outputs are exposed to control rApps/xApps implementing energy efficient automated control-loops. As an exemplary use case, we described the required workflows to implement an AI-driven cell on/off control. Future work includes the definition and implementation of the AI Engine Assist rApp/xApp model and SDK to support the AI/ML workflows. We also plan to further elaborate and integrate uses cases comprehending the utilization of the O1+ and O2+ interfaces. 


\section*{Acknowledgment}
This work is supported by the Smart Networks and Services Joint Undertaking (SNS JU) under the European Union’s Horizon Europe research and innovation programme under Grant Agreement No 101097083, BeGREEN project. Views and opinions expressed are however those of the author(s) only and do not necessarily reflect those of the European Union or SNS-JU. Neither the European Union nor the granting authority can be held responsible for them. 

\bibliographystyle{IEEEtran}
\bibliography{References}



%

\end{document}